\documentclass[sigconf,usenames,dvipsnames,table]{acmart}
\usepackage{comment}
\usepackage{graphicx}
\usepackage{url}
\usepackage[utf8]{inputenc}
\usepackage{endnotes}
\usepackage{comment}
\usepackage{graphicx}
\usepackage{url}
\usepackage{color}
\usepackage{listings}
\usepackage{float}
\usepackage{lipsum}
\usepackage{enumitem}
\usepackage{tabularx}
\usepackage{makecell}
\usepackage{tikz}
\usetikzlibrary{backgrounds,fit, decorations.markings, patterns}
\usepackage{pgfplots}
\pgfplotsset{compat=1.5}
\usepackage{amssymb,amsmath}
\usepackage{ifthen}
\usepackage{soul}
\usepackage{seqsplit}
\usepackage{xspace}
\usepackage{framed}
\usepackage{amsthm}
\usepackage{subcaption}
\usepackage{listings}
\usepackage{courier}

\newboolean{showcomments}

\setboolean{showcomments}{true}

\ifthenelse{\boolean{showcomments}}
  {\newcommand{\nb}[2]{
    \fbox{\bfseries\sffamily\scriptsize#1}
    {\sf\small$\blacktriangleright$\textit{#2}$\blacktriangleleft$}
   }
   
  }
  {\newcommand{\nb}[2]{}
   
  }

\newcommand\finding[1]{\vspace{0.25em}\noindent\textsf{\bf Finding {#1}.}}
\newcommand\fnumber[1]{{$\mathcal{F}_{#1}$}}
\newcommand\myparagraph[1]{\vspace{0.25em}\noindent{\bf {#1}:}}

\newcommand{\nest}{{Nest}\xspace}
\newcommand{\review}{{product review}\xspace}
\newcommand{\hue}{{Hue}\xspace}
\newcommand{\nestapps}{Apps$_{\small nest}$\xspace}	
\newcommand{\gplayapps}{Apps$_{\small general}$\xspace}	
\newcommand{\nestappsext}{Apps$_{\small nestExt}$\xspace}

\newcommand{\ie}{\textit{i.e.,}\xspace}
\newcommand{\eg}{\textit{e.g.,}\xspace}

\definecolor{darkgreen}{RGB}{0,102,0}
\definecolor{darkorange}{RGB}{255, 102, 0}

\lstdefinestyle{javaStyle} {
  language=Java,
  showspaces=false,
  showtabs=false,
  breaklines=false,
  showstringspaces=false,
  breakatwhitespace=true,
  numbers=left,
  numberstyle=\scriptsize,
  tabsize=2,
  captionpos=b,
  commentstyle=\bfseries\color{gray},
  keywordstyle=\bfseries\color{Plum},
  stringstyle=\color{red}\bfseries,
  basicstyle=\ttfamily\footnotesize,
  moredelim=[il][\textcolor{lightgray}]{\$\$},
  moredelim=[is][\textcolor{lightgray}]{\%\%}{\%\%}
}

\acmDOI{10.475/123_4}

\acmISBN{123-4567-24-567/08/06}

\setcopyright{none}
\acmConference[CODASPY 2019]{ACM Conference on
Data and Application Security and Privacy}{March 2019}{Dallas, TX, USA}
\acmYear{2019}
\copyrightyear{2019}

\title{A Study of Data Store-based Home Automation}
\author{{\rm Kaushal Kafle, Kevin Moran, Sunil Manandhar, Adwait Nadkarni,
Denys Poshyvanyk}}
\affiliation{William \& Mary, Williamsburg, VA, USA}
\email{{kkafle, kpmoran, smanandhar, nadkarni, denys}@cs.wm.edu}
\date{}

\begin{document}

\renewcommand{\shortauthors}{K. Kafle, K.Moran, S. Manandhar, A. Nadkarni, and D. Poshyvanyk}

\begin{abstract}

Home automation platforms provide a new level of convenience by enabling
consumers to automate various aspects of physical objects in their
homes.
While the convenience is beneficial, security flaws in the platforms or
integrated third-party products can have serious consequences for the
integrity of a user's physical environment. 
In this paper we perform a systematic security evaluation of two popular
smart home platforms, Google's \nest platform and Philips \hue, that
implement home automation ``routines'' (\ie trigger-action programs
involving apps and devices) via manipulation of state variables in a
{\em centralized data store}.  Our semi-automated analysis examines,
among other things, platform access control enforcement, the rigor of
non-system enforcement procedures, and the potential for misuse of
routines. 
This analysis results in {\em ten} key findings with serious security
implications. For instance, we demonstrate the potential for the misuse
of smart home routines in the Nest platform to perform a lateral
privilege escalation, illustrate how \nest's \review system is
ineffective at preventing multiple stages of this attack that it
examines, and demonstrate how emerging platforms may fail to provide
even bare-minimum security by allowing apps to arbitrarily add/remove
other apps from the user's smart home.
Our findings draw attention to the unique security
challenges of platforms that execute routines via centralized data
stores, and highlight the importance of enforcing security by design in
emerging home automation platforms.
\end{abstract}


\maketitle

\section{Introduction}
\label{sec:intro}

Internet-connected, embedded computing objects known as \textit{smart
home products} have become extremely popular with consumers. The utility
and practicality afforded by these devices has spurred tremendous market
interest, with over 20 billion smart home products projected to be in
use by 2020~\cite{iot-adoption}. The diversity of these products is
staggering, ranging from small physical devices with
embedded computers such as smart locks and light bulbs, to full fledged
appliances such as refrigerators and HVAC systems.  In the modern
computing landscape, smart home devices are unique as they provide an
often imperceptible bridge between the digital and physical worlds by
connecting physical objects to digital services via the Internet,
allowing the user to conveniently automate their home.  However,
because many of these products are tied to the user's security or
privacy (\eg door locks, cameras), it is important to understand the
attack surface of such devices and platforms, in order to
build practical defenses without sacrificing utility.

As the market for smart home devices has continued to mature, a new
software paradigm has emerged to facilitate smart home automation via
the interactions between smart home devices and the apps that control
them.    These interactions may be expressed as {\em routines}, which
are sequences of app and device actions that are executed upon one or
more triggers, \ie an instance of the trigger-action paradigm in the
smart home. Routines are becoming the foundational unit of home
automation~\cite{routinegoogle1,routinealexa,routinesmartthings,routineall},
and as a result, it is natural to characterize existing platforms based
on how routines are implemented.  

If we categorize available platforms based on how routines are
facilitated, we observe two broad categories: {\sf (1)} {\sf API-based
Smart Home Managers} such as Yeti~\cite{yetiapp},
Yonomi~\cite{yonomiapp}, IFTTT~\cite{iftttapp}, and
Stringify~\cite{stringifyapp} that allow users to chain together a
diverse set of devices using third-party APIs exposed by device vendors,
and {\sf (2)} smart home platforms such as Google's Works with
\nest~\cite{nestdoc}, Samsung SmartThings~\cite{smartthings}, and
Philips \hue~\cite{philipshue} that leverage {\em centralized data
stores} to monitor and maintain the states of IoT devices. We term these
platforms as {\sf Data Store-Based (DSB) Smart Home Platforms}. In DSB
platforms, complex routines are executed via reads/writes to state
variables in a central data store.

This paper is motivated by a key observation that while routines are
supported via centralized data stores in all DSB platforms, there are
differences in the manner in which routines are created, observed, and
managed by the user. That is, SmartThings encourages users to take full
control of creating and managing routines involving third-party apps and
devices via the SmartThings app. On the contrary, in \nest, users do not
have a centralized perspective of routines at all, and instead, manage
routines using third-party apps/devices. This key difference may imply
unique security challenges for \nest. Similarly, being a much simpler
platform within this category of DSB platforms, \hue represents another
unique and interesting instance of the DSB platform paradigm.  While
prior work has explored the security of routines enabled by a smart home
manager (\ie specifically, IFTTT recipes~\cite{sab+17}), the permission
enforcement and application security in the SmartThings
platform~\cite{fjp16}, 
and the side-effects of SmartThings SmartApps~\cite{cmt18},
there is a notable gap in current research. 
Namely, prior studies do not evaluate the potential for {\em
adversarial} misuse of routines, which are the essence
of DSB platforms, and by extension, home automation. 

\myparagraph{Contributions}
This paper performs a systematic security analysis of some of the less
studied, but widely popular, data store-based smart home platforms,
\ie~\nest and \hue, helping to close the existing gap in prior research.
In particular, we evaluate {\sf (1)} the access control enforcement in
the platforms themselves, {\sf (2)} the robustness of other non-system
enforcement (\eg product reviews in \nest), {\sf (3)} the use and more
importantly the {\em misuse} of routines via manipulation of the data
store by low-integrity devices,
\footnote{In the context of our study, we define a device as
high-integrity if it is advertised as security-critical by the device
vendor (\eg ~ \nest Cam) while those that are not security-critical are
referred to as low-integrity (\eg Philips Hue lamp).} 
and finally, {\sf (4)} the security of applications that integrate into
these platforms. 
To our knowledge, this paper is the  first to analyze this relatively
new class of smart home platforms, in particular the \nest and \hue
platforms, and to provide a holistic analysis of routines, their use,
and potential for their misuse in DSB platforms. Moreover, this paper is
the first to analyze the accuracy of app-defined permission prompts,
which form one of the few sources of access control information for the
user.  Our novel findings (\fnumber{1}$\rightarrow$\fnumber{10}),
summarized as follows, demonstrate the unique security challenges faced
by DSB platforms at the cost of seamless automation:

\vspace{-0.1cm}
\begin{itemize}[leftmargin=1em]
  \item {\bf Misuse of routines} -- The permission model in \nest is
    fine-grained and enforced according to specifications (\fnumber{1}),
    giving low-integrity third-party apps/devices (\eg a switch) little
    room for directly modifying the data store variables of
    high-integrity devices (\eg security cameras). However, the routines
    supported by \nest allow low-integrity devices/apps to indirectly
    modify the state of high-integrity devices, by modifying the shared
    variables that both high/low integrity devices rely on
    (\fnumber{4}).  
  \item {\bf Lack of systematic defenses} --
    \nest does not employ transitive access control enforcement to
    prevent indirect modification of security-sensitive data store
    variables; instead, it relies on a \review of application artifacts
    before allowing API access.  We discover that the \review process is
    insufficient and may not prevent malicious exploitation of routines;
    \ie the review mandates that apps prompt the user before modifying
    certain variables, but does not validate {\em what} the prompt
    contain, allowing apps to deceive users into providing consent
    (\fnumber{5}).  Moreover, permission descriptions provided by apps
    during authorization are also often incorrect or misleading
    (\fnumber{6}, \fnumber{9}), which demonstrates that malicious apps
    may easily find ways to gain more privilege than necessary
    (\fnumber{7}), circumventing both users and the \nest\ \review
    (\fnumber{8}). 
  \item {\bf Lateral privilege escalation} -- We find that smart home
    apps, particularly those that connect to \nest and have permissions
    to access security-sensitive data store variables, have a
    significantly high rate of SSL vulnerabilities (\fnumber{10}).  We
    combine these SSL flaws with the findings discussed previously
    (specifically \fnumber{4}$\rightarrow$\fnumber{9}) and demonstrate a
    novel form of a {\em lateral} privilege escalation attack. That is, we
    compromise a low-integrity app that has access to the user's \nest
    smart home (\eg a TP Link Kasa switch), use the
    compromised app to change the state of the data store
    to trigger a security-sensitive routine, and indirectly
    change the state of a high-integrity \nest device (\eg the \nest
    security camera).  This attack can be used to deceive the \nest Cam
    into determining that the user is home when they are actually away,
    and prevent it from monitoring the home in the user's absence.
  \item {\bf Lack of bare minimum protections} -- Unlike \nest, the
    access control enforcement of \hue is woefully inadequate.  Third-party 
    apps that have been added to a user's \hue platform may
    arbitrarily add other third-party apps without user consent, despite 
	an existing policy that the user must consent by physically
    pressing a button (\fnumber{2}).  Making matters worse, an app may
    {\em remove} other apps integrated with the platform by exploiting
    unprotected data store variables in \hue (\fnumber{3}). These
    vulnerabilities may allow an app with seemingly useful functionality
    (\ie a Trojan~\cite{lbmc94}) to install malicious add-ons in a
    manner invisible to the user, and replace the user's integrated apps
    with its malicious substitutes.   
\end{itemize}

The rest of the paper is structured as follows:
Section~\ref{sec:characteristics} describes the key attributes of DSB
platforms, and provides background.  Section~\ref{sec:overview} provides
an overview of our analysis, and
Sections~\ref{sec:permission-map},~\ref{sec:platform-routines}
and~\ref{sec:app} describe our analysis of permission enforcement in
\nest and \hue, security ramifications of routines, and security of
smart home apps, respectively.
Section~\ref{sec:attack} provides an end-to-end attack, and
Section~\ref{sec:lessons} describes the lessons learned.
Section~\ref{sec:reporting} describes the vendors' response to our
findings.  Section~\ref{sec:threats} describes the threats to validity.
Section~\ref{sec:relwork} describes the related work, and
Section~\ref{sec:conc} concludes. 

\section{Home Automation via Centralized Data Stores}
\label{sec:characteristics}
 
This section describes the general characteristics of data
store-based platforms, \ie smart home platforms that use a
\textit{centralized data store} to facilitate routines. Following this
general description, we provide background on two such platforms, namely
{\sf (1)} Google's ``Works with~\nest''~\cite{nest} platform (henceforth
called ``\nest'') and {\sf (2)} the Philips~\hue lighting
system~\cite{hue} (henceforth called ``\hue''), which serve as the
targets of our security analysis. While there are no official statistics
on the market adoption of either \nest or \hue, the Android apps for
both of the systems have over a million downloads on Google
Play~\cite{neststats,huestats}, indicating significant adoption, and
far-reaching security impact of our analysis.

\begin{figure}[t]
    \centering
    \includegraphics[width=3in]{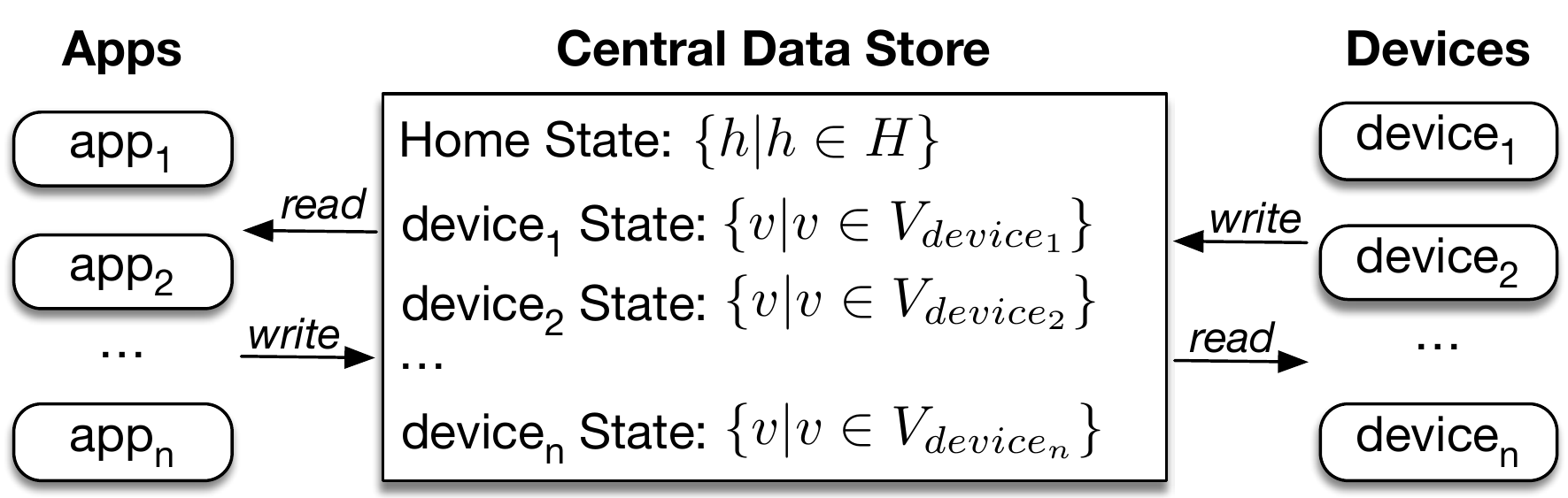} 
    \vspace{-0.3cm}
    \caption{{\small The general architecture of home automation
platforms that leverage centralized data stores. Note that $H$ is the universe
of all home state variables, and $V_{device_i}$ is the universe of all state
variables specific to {\em device$_i$}}}
\label{fig:data_store_platform}
\vspace{-1em}
\end{figure} 

\vspace{-0.8em}
\subsection{General Characteristics}
\vspace{-0.2em}
Figure~\ref{fig:data_store_platform} describes the general architecture
of DSB platforms, consisting of
three main components: \textit{apps}, \textit{devices}, and the
\textit{centralized data store}. These components generally communicate
over the Internet. Additionally, a physical hub that facilitates local
communication via protocols such as Zigbee or Z-wave may or may not be
included in this setup (\eg the \hue Bridge); \ie in a general sense,
routines are agnostic of the presence of the hub. Hence, we exclude the
hub in Figure~\ref{fig:data_store_platform}. Similarly, the apps may
either be Web services hosted on the cloud, or mobile apps communicating
via Web services. At this juncture, we generalize apps as third-party
software interacting with the data store, and provide the specifics for
individual platforms in later sections.

The centralized data store facilitates communication among apps and
devices via state variables.  The data store exposes two types of state
variables: {\sf (1)} {\em Home} state variables that reflect the general
state of the entire smart home (\eg~if the user is at {\em home/away},
the {\em devices attached} to the home, the {\em postal code}), and {\sf
(2)} {\em Device-specific} state variables that reflect the attributes
specific to particular devices (\eg~if the Camera is {\em streaming},
the {\em target temperature} of the thermostat, the {\em battery health
} of the smoke alarm).  

Apps and devices communicate by reading from or writing to the state
variables in the centralized data store. This model allows expressive
communication, from simple state updates to indirect trigger-action
routines. Consider this simple state update: the user may change the
temperature of the thermostat from an app, which in turn {\em writes}
the change to the {\em target temperature} variable in the data
store.  The thermostat device receives an update from the data store
(\ie ~{\em  reads} the {\em target temperature} state variable), and
changes its target temperature accordingly. Further, as stated
previously, expressive routines may also be implemented using the data
store. For instance, the thermostat may change to its ``economy'' mode when
the home's state changes to {\em away}.  That is, the thermostat's app
may detect that the user has left the smart home (\eg using Geofencing),
and {\em write} to the home state variable {\em away}. The thermostat
may then {\em read} this change, and switch to its economy mode. 

A salient characteristic of DSB platforms is that they lean towards
seamless home automation, by automatically interacting with devices and
executing complex routines via the centralized data store. 
However, even within platforms that follow this model (\eg Samsung
SmartThings, \nest, and \hue), our preliminary
investigation led to the following {\em key observations} that motivate
a targeted analysis of the \nest and \hue platforms and their apps: 

\myparagraph{Key Observations}
We observe that both \nest and SmartThings execute routines; however,
there is a key difference in how routines are managed.
SmartThings allows users to create and manage routines from the
SmartThings app itself, thereby providing users with a general view of
all the routines executing in the home~\cite{smartthingsroutines}. In
contrast, \nest routines are generally implemented as {\em decentralized} third-party
integrations. Third-party products that
facilitate routines provide the user with the ability to
view and manage them. As a result, the \nest platform does not provide the
user with a {\em centralized view} of the routines that are in
place. Due to this lack of user control, \nest smart homes may face
unique security risks and challenges, which motivates this security
analysis. Similarly, we observe that the Philips \hue platform may be
another interesting variant of DSB platforms.  That is, \hue integrates
{\em homogeneous} devices related to lighting such as lamps and bulbs,
unlike \nest and SmartThings that integrate heterogeneous devices, and
represents a drastically simpler (and hence unique) variant of home
automation platforms that use centralized data stores. As a result, the
analysis of \hue's attack surface has potential to draw attention to
other similar, homogeneous platforms, which is especially important
considering the fragmentation in the smart home product
ecosystem~\cite{smarthome-frag}.  To our knowledge, this paper is the
first to analyze this relatively new class of smart home platforms, and
specifically, \nest and \hue.

\vspace{-1em}
\subsection{\nest Background}
\label{sec:nest}

\begin{figure}[t]
    \centering
    \includegraphics[width=2.4in]{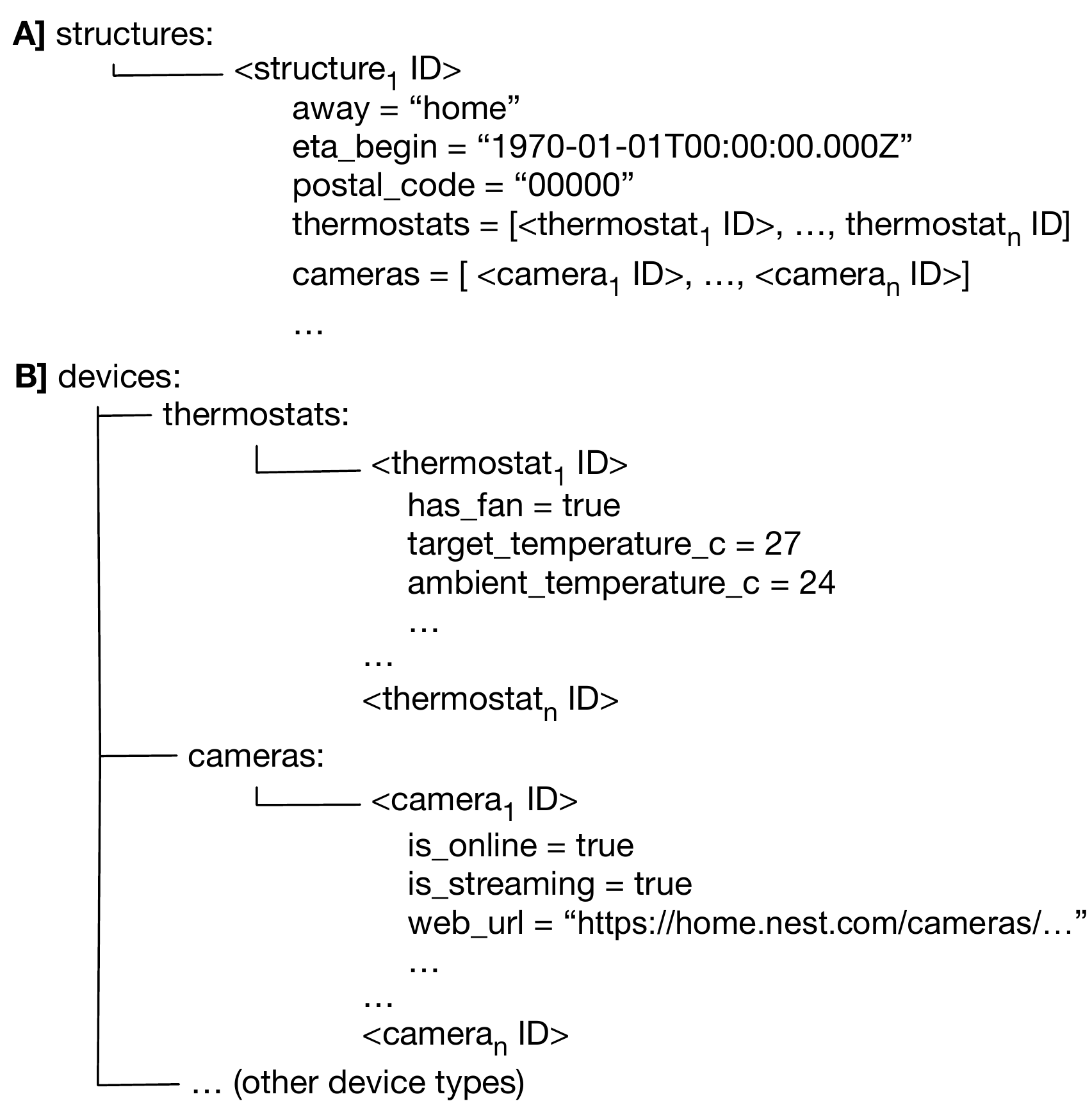} 
    \vspace{-0.7em}
    \caption{{\small A simplified view of the centralized data store in \nest.
}}
\label{fig:nest_data_store}
\vspace{-1.3em}
\end{figure} 

The \textit{Works with \nest} platform integrates a heterogenous set of devices,
including devices from \nest (\eg Nest thermostat, \nest Cam, \nest
Protect) as well as from other brands (\eg Wemo and Kasa switches,
Google Home, MyQ Chamberlain garage door opener)~\cite{nest}.
This section describes the key characteristics of \nest, \ie its data
store, its access control model, and routines.

\myparagraph{Data store composition} Figure~\ref{fig:nest_data_store}
shows a simplified, conceptual view of the centralized data store in
\nest. Note that the figure shows a small fraction of the true data
store, \ie only enough to facilitate understanding. \nest implements the
data store as a JSON-format document divided into two main top-level
sections: {\em structures} and {\em devices}. A {\em structure}
represents an entire smart home environment such as a user's home or
office, and is defined by various state variables that are global across
the smart home (\eg~{\em Away} to indicate the presence or absence of
the user in the structure and the {\em postal\_code} to indicate the
home's physical location).  The devices are subdivided into device types
(\eg thermostats, cameras, smoke detectors), and there can be many
devices of a certain type, as shown in Figure~\ref{fig:nest_data_store}.
Each device stores its state in variables that are relevant to its type;
\eg a thermostat has state variables for {\em humidity}, and {\em
target\_temperature\_c}, whereas a camera has the variables {\em
is\_online} and {\em is\_streaming}.  Aside from these type-specific
variables, devices also have certain variables in common; \eg the
alphanumeric {\em device ID}, the {\em structure ID} of the structure in
which the device is installed, the device's user-assigned {\em name},
and {\em battery\_health}. 

\myparagraph{Access Control in \nest.} 
\nest treats third-party apps, Web services, and devices that want to
integrate with a \nest-based smart home as ``products''. Each \nest user
account has a specific data store assigned to it and any product that
requests access to the user's data store needs to be first authorized by
the user using OAuth 2.0.  \nest defines read or read/write permissions
for each of the variables in the data store.  Additionally, some
variables {\eg the list of all thermostats in the structure} are always
{\em read-only}. A product that wants to register with \nest must first
declare the permissions that it needs (\eg {\em thermostat read}, {\em
thermostat read/write}) in the \nest developer console.  When connecting
a product to \nest, during the OAuth authorization phase, the user is
shown the permissions requested by the product. Once the user grants the
permissions, a revocable access token is generated specific to the
product, the set of permissions requested, and the particular smart home
to which the product is connected. This token is used for subsequent
interactions with the data store.

\myparagraph{Accessing the \nest data store.} Devices and applications
that are connected to a particular smart home (\ie the user's \nest
account) can update data store variables to which they have access, and
also subscribe to the changes to the state of the data store. Nest uses
the REST approach for these update communications, as well as for
apps/devices to modify the data store. The REST endpoints can be
accessed through HTTPS by any registered Nest products.  

\myparagraph{Routines in \nest} The ability of connected devices to
observe and write to state variables in the centralized data store
facilitates trigger-action routines. However, in \nest, the user cannot
create or view routines in a centralized interface (\ie unlike
SmartThings). Instead, apps may provide routines as opt-in features.
For example, the \nest smoke alarm's {\em smoke\_alarm\_state} variable has three possible values, ``ok'', ``warning'', and ``emergency''.
When this variable is changed to ``warning'', other smart home products
(e.g., Somfy Protect~\cite{somfyprotectwwn}) can be
configured to trigger and warn the user.  Note that in the {\em
Home/Away assist} section of the \nest app settings, users can view
a summary of how certain variables (i.e., home or away) affect their
Nest-manufactured devices; however, there is no way for users to observe the
triggers/apps that change the state of the {\em away} variable {\em
simultaneously} with the resultant actions, preventing them from fully
understanding how routines execute in their home.

\vspace{-1.5em}
\subsection{\hue Background}
\label{sec:hue}

Unlike \nest, which is a platform for heterogeneous devices, Philips
\hue deals exclusively with lighting devices such as lamps and bulbs. As
a result, the centralized data store of Philips Hue supports much
simpler routines.  \hue implements its data store as a JSON document
with sections related to {\sf (1)} physical lighting devices, {\sf (2)}
semantic groups of these devices, and {\sf (3)} global config variables
(such as \textit{whitelisted apps} and the \textit{linkbutton}). To
connect a third-party management app to a user's existing \hue system,
the app identifies a \hue bridge connected to the local network, and
requires the user to press a physical button on the bridge. Once this
action is completed by the user, the app receives a \textit{username}
token that is stored in the \textit{whitelisted} section of the \hue
data store.  Whitelisted apps can then read and modify data store
variables as dictated by \hue's access control policy, which grants all
authorized apps the same access regardless of their purported
functionality.  Our online appendix provides additional details regarding the \hue platform~\cite{online_appendix}.

\section{Analysis Overview}
\label{sec:overview}
This paper analyzes the security of home automation platforms that rely
on centralized data stores (\ie DSB platforms). Third-party apps are the
security principals on such platforms, as they are assigned specific
permissions to interact with the integrated devices. That
is, as described in Section~\ref{sec:characteristics}, DSB platforms
consist of (1) {\em third-party apps} that interact with the smart home
(\ie centralized data store and devices) by acquiring (2) {\em platform
permissions}, and execute a complex set of such interactions as (3) {\em
trigger-action routines}. Our analysis methodology takes these three
aspects into consideration, starting with platform permissions, as
follows:

\myparagraph{A. Analysis of Platform Permissions
(Section~\ref{sec:permission-map})} We analyze the {\em enforcement} of
platform permissions/access control to discover inconsistencies. For
this analysis, we automatically build permission maps, and
semi-automatically analyze them.

\myparagraph{B. Analysis of Routines
(Section~\ref{sec:platform-routines})} While analyzing permission
enforcement gives us an idea of what individual devices can accomplish
with a certain set of permissions, we perform an experimental analysis
with real devices to identify the interdependencies among devices
and apps through the shared data model, and the ramifications of such
interdependencies on the user's security and privacy. Additionally, we
notice that \nest does not enforce transitive access control policies to
prevent dangerous side-effects of routines, but instead employs a
product review process as a defense mechanism. We analyze the
effectiveness of this review process using the permission prompts used
by existing apps as evidence.

\myparagraph{C. Analysis of Third-party Apps (Section~\ref{sec:app})} We
analyze the permission descriptions presented by mobile apps compatible
with \nest to identify over-privileged apps, or apps whose
permission descriptions are inconsistent with the permission requested.
We then analyze the apps for signs of SSL misuse, in order to exploit
applications that possess critical permissions, which can be leveraged
to indirectly exploit security critical devices in the smart home.

We combine the findings from these three analyses to demonstrate an
instance of a {\em lateral privilege escalation} attack in a smart home
(Section~\ref{sec:attack}). That is, we demonstrate how an attacker can
compromise a low-integrity device/app integrated into a smart home (\eg
a light bulb), and use routines to perform protected operations on a
high-integrity product (e.g., a security camera).

\section{Evaluating Permission Enforcement}
\label{sec:permission-map}

The centralized data store described in
Section~\ref{sec:characteristics} may contain variables whose secrecy or
integrity is crucial; \eg unprotected write access to the {\em web\_url}
field of the camera may allow a malicious app to launch a phishing
attack, by replacing the URL in the field with an attacker-controlled
one. To understand if appropriate barriers are in place to protect such
sensitive variables, we perform an analysis of the permission
enforcement in \nest and \hue.

Our approach is to generate and analyze the {\em permission map} for
each platform, \ie the variables that can be accessed with each
permission, and inversely, the permissions needed to access each
variable of the data store. Note that while this information should
ideally be available in the platform documentation, prior analysis of
similar systems has demonstrated that the documentation may not always
be complete or correct in this regard~\cite{fch+11,fjp16}.

\vspace{-0.9em}
\subsection{Generating Permission Maps}
\vspace{-0.1em}

We generate the permission map using automated testing as in prior work on Android~\cite{fch+11}. 
We use two separate approaches for \nest and \hue, owing to their
disparate access control models.

\myparagraph{Approach for \nest}
We first created a simulated home environment using the \nest Home
Simulator~\cite{nestsimulator}, and linked our \nest user account to
this simulated smart home. We then created our test Android app, and
connected our test app to the simulated home (\ie our \nest user
account) as described in Section~\ref{sec:nest}. Note that the simulated
smart home is virtually identical to an end-user's setup, such that real
devices may be added to it. Using the simulator allows us to investigate
the data store information of \nest devices (\eg the Smoke/CO detector)
that we may not have installed.  

In order to generate a complete view of the data store, we granted our
test app all of the 15 permissions in \nest (\eg {\em Away read/write},
{\em Thermostat read}), and read all accompanying information. To build
the permission map for \nest's 15 permissions, we created 15 apps, such
that each app requested a single unique permission, and registered these
apps to our developer account in the \nest developer console. Note that
we do not test the effect of permission combinations, as our goal is to
test the enforcement of individual permissions, and \nest's simple
authorization logic simply provides an app with a union of the
privileges of the individual permissions.

We then connected each of the 15 apps to our \nest user account using
the procedure described in 
Section~\ref{sec:nest}. We programmed each app to attempt to read
and write each variable of the data store (\ie the
previously derived {\em complete view}).
We recorded the outcome of each access, \ie if it was successful, or an
access control denial. In the cases where we experienced non-security
errors writing to data store variables (\eg writing data with an
incorrect type), we revised our apps and repeated the test. The outcome
of this process was a permission map, \ie the mapping of each permission
to the data store variables that it can read and/or write.

\myparagraph{Approach for \hue}
We followed the procedure for \hue described in
Section~\ref{sec:hue} to get a unique token that registers our
single test app with the data store of our \hue bridge.  
In \hue, all the variables of the data store are ``readable'' (\ie we
verified that all the variables described in the developer
documentation~\cite{philipshue} can be read by third-party apps).  Therefore, to build
the permission map, we first extracted the contents of the entire data
store. Then, for each subsection within the data store, our app made
repeated write requests, \ie PUT calls with the payload consisting of a
dummy value based on the variable type (\ie String, Boolean and Integer).
All the variables that were successfully written to using this
method were assigned as ``writable'' variables. Similarly, our app made
repeated DELETE calls to the API and the variables that were
successfully deleted were assigned as ``writable'' variables.
This generated permission map applies to all third-party apps connected
to \hue, since the platform provides equal privilege to all third-party apps.

\vspace{-0.9em}
\subsection{Analyzing Permission Maps} 
\vspace{-0.1em}

The objective behind obtaining the permission map is to understand the
potential for application overprivilege,  by analyzing the granularity
as well as the correctness of the enforcement. We analyze the permission
map to identify instances of {\sf (1)} {\em coarse-grained permissions},
\ie permissions that give the third-party app access to a set of
security-sensitive resources that must ideally be protected under
separate permissions, and {\sf (2)} {\em incorrect enforcement}, \ie
when an app has access to more resources (\ie state variables) than it
should have given its permission set, as per the documentation; \eg apps
on SmartThings may lock/unlock the door lock without the explicit
permission required to do so~\cite{fjp16}.

To perform this analysis, we first identified data store variables that
may be security or privacy-sensitive. This identification was performed
using an open-coding methodology by one author, and separately verified
by another author, for each platform.
We then performed further analysis by separately considering each such
variable, and the permission(s) that allow access to it.  A major
consideration in our analysis is the security impact of an adversary
being allowed read or read/write access to a particular resource.
Moreover, our evaluation of the impact of the access control enforcement
was contextualized to the platform under inspection. That is, when
evaluating \nest, we took into consideration the semantic meaning and
purpose of certain permissions in terms of the data store variables, as
described in the documentation (\eg that the {\em Away read/write}
permission should be required to write to the {\em away}
variable~\cite{nestperms}). For \hue, we only considered the
security-impact of an adversary accessing data store variables. Our
rationale is that the \hue platform defines the same static policy (\ie
same permissions) for all third-party apps, and hence, its permission
map can be simply said to consist of just one permission that provides
access to a fixed set of data store variables.  As a result, we judge
application over-privilege in \hue by considering the impact of an
adversarial third-party app reading from or writing to each of the
security-sensitive variables identified in \hue's permission map.

The creation of the permission maps for both \nest and \hue requires the
application of well-studied automated testing techniques, and as such,
can be replicated for similar platforms, with minor changes to input
data (\eg the permissions to test for). We will release our
code and data to developers and platform vendors.  

\vspace{-0.2cm}
\subsection{Permission Enforcement
Findings~{\normalsize($\mathcal{F}_1\rightarrow\mathcal{F}_3$)}}
\vspace{-0.1cm}

\finding{1: The permission enforcement in \nest is fine-grained and
correctly enforced, \ie as per the specification ($\mathcal{F}_1$)} We
observe that the \nest permission map is significantly  more
fine-grained, and permissions are correctly enforced, relative to the
observations of prior research in similar platforms (\eg the analysis of
SmartThings~\cite{fjp16}). Some highly sensitive variables are always
read-only (\eg the {\em web\_url} where the camera feed is posted), and
there are separate read and read/write permissions to access sensitive
variables. Variables that control the state of the entire smart home
are protected by dedicated permissions that control write privilege; \eg
the {\em away} variable can only be written to using the {\em Away
read/write} permission, the {\em ETA} variable has separate permissions
for apps to read and write to it (\ie~ {\em ETA read} and {\em ETA
write}), and the Nest Cam can only be turned on/off via the {\em
is\_streaming} variable, using
the{\em~Camera + Images read/write} permission that controls write
access to it. Moreover, since many apps need to
respond to the {\em away} variable (\ie react when the user is
home/away), device-specific read permissions (\eg~{\em Thermostat read},
{\em Smoke + CO read}) also allow apps to {\em read} the {\em away}
variable, eliminating the need for apps to ask for higher-privileged
{\em Away read} permission.  The separate read and read/write
permissions are correctly enforced, \ie {\em our generated permission
map provides the same access as is defined in the \nest permission
documentation~\cite{nestperms}}. This is in contrast with findings of
similar analyses of permission models in the past (\eg the Android
permission model~\cite{fch+11}, SmartThings~\cite{fjp16}), and
demonstrates  that the \nest platform has incorporated lessons from
prior work in permission enforcement.  

\finding{2: In \hue, the access control policy allows apps to bypass the
user's explicit consent ($\mathcal{F}_2$)} 
We discovered two data store variables that were not write-protected,
and which have a significant part to play in controlling access to the
data store and the user's smart home.  First, any third-party app
can write to the {\em linkbutton} flag.  Recall from
Section~\ref{sec:hue} that the user has to press the physical button on
the \hue bridge device to authorize an app's addition to the bridge. The
physical button press changes the {\em linkbutton} value to ``true'',
and allows the app to be added to the {\em whitelist} of allowed
third-party apps.  However, we discovered that once installed, an app
can toggle the {\em linkbutton} variable at will, {\em enabling third-party
apps to add other third-party apps to the smart home without the
user's consent}. This exploitable access control vulnerability can allow
an app with seemingly useful functionality to install malicious add-ons
by bypassing the user altogether. In our tests, we verified this attack
with apps that were connected to the local network. This condition is
feasible as a malicious app that needs to be added without the user's
consent may not even have to pretend to work with \hue; all it needs is
to be connected to the local network (\ie a game on the
mobile device from one of the people present in the smart home). Note that
it is also possible to remotely perform this attack, which we discuss in
Section~\ref{sec:threats}.

\finding{3. In \hue, third-party apps can directly modify the list of
added apps, adding and revoking access without user consent
($\mathcal{F}_3$)}  \hue stores the authorization tokens of apps
connected to the particular smart home in a {\em whitelist} on the \hue
Bridge device.  While analyzing the permission map, we discovered that
not only could our third-party test app read from this list, it could
also directly delete tokens from it. We experimentally confirmed this
finding again, by removing {\em Alexa} and {\em Google Home} from the
smart home, without the user's consent. An adversary could easily
combine this vulnerability with ($\mathcal{F}_2$), to remove legitimate
apps added by the user, add adversary-controlled apps (\ie by keeping
the {\em linkbutton} ``true''), all without the user's consent. More
importantly, users do not get alerts when such changes are made (\ie
since it is assumed that the enforcement will correctly acquire user
consent). Hence, unless the user actually checks the list of integrated
apps using the \hue Web app, the user would not notice these changes.

While the \nest permission model is robust in its mapping of data store
variables and permissions required to access them,
Section~\ref{sec:platform-routines} demonstrates how fields
disallowed by permissions may be indirectly modified via strategic
misuse of routines, and describes \nest's product review guidelines to
prevent the same~\cite{nestreview}.  Section~\ref{sec:app} describes how
badly written and overprivileged apps escape these review guidelines,
and motivate a technical solution.

\section{Evaluating Smart Home Routines}
\label{sec:platform-routines}

Prior work has demonstrated that in platforms that favor application
interoperability but lack transitive access control enforcement,
problems such as confused deputy and application collusion may
persist~\cite{fwm+11,cfgw11,ne13,naej16}. Smart homes that
facilitate routines are no different, but the exploitability and impact
of routines on smart homes is unknown, which motivates this aspect of
our study.

Recall that routines are trigger-action programs that are either
triggered by a change in some variable of the data store, or whose
action modifies certain variables of the data store. While both \nest
and \hue share this characteristic, routines in \hue are fairly limited
in scope, and their exploitation is bound to only affect the lighting of
the smart home. As a result, while we provide confirmed
examples of \hue routines in Section~\ref{sec:hue}, the security
evaluation described in this section is focused on the heterogeneous
\nest platform that facilitates more diverse and expressive routines.

\vspace{-0.9em}
\subsection{Methodology for the Analysis of Routines} 
\vspace{-0.1em}
While using the simulator as described in Section~\ref{sec:permission-map} allows us to understand what routines are {\em
possible} on the platform, \ie what variables might be manipulated, and
what \nest devices (e.g., the \nest Cam, \nest Thermostat) are affected
as a result, we performed additional experiments with real apps and
devices to study existing routines in the wild. For this experiment, we
extended the smart home setup previously discussed in
Section~\ref{sec:permission-map} with real devices.

We started by collecting a list of devices that integrate with \nest
from the \textit{Works with \nest} website~\cite{nest}. Using this initial list and
information from the website, we purchased a set of 7 devices that
possessed a set of characteristics relevant to this study, \ie devices
that {\sf (1)} take part in routines (\ie as advertised on the website),
{\sf (2)} are important for the user's security or privacy, and {\sf
(3)} are widely-known/popular with a large user base (\ie determined by
the number of installs of the mobile client on Google Play). We obtained
a final list of devices (7 real and 2 simulated) to our \nest smart
home, namely, the \nest Cam (\ie a security camera), \hue light bulb,
Belkin Wemo switch, the MyQ Chamberlain garage door opener, TP Link Kasa
Smart Plug, Google Home, Alexa, \nest Thermostat (simulated), and the
\nest Protect Smoke \& CO Alarm (simulated).  Some devices that may be
important for security did not participate in routines at the time of
the study, and hence were excluded from our final device list. 

We connected these devices to our \nest smart home using the Android
apps provided by device vendors, and connected a small set of smart
home managers (\eg Yeti~\cite{yetiapp} and Yonomi~\cite{yonomiapp}) to
our \nest smart home as well.  For each device, we set up and executed
each individual routine as described on the Works with Nest as
well as the device vendor's website, and observed the effects on the
rest of the smart home (especially, security-sensitive devices). 
Also, we manipulated data store variables
from our test app, and observed the effects on previously configured
routines and devices.

\vspace{-0.2cm}
\subsection{Smart Home Routine Findings {\normalsize($\mathcal{F}_4 \rightarrow
\mathcal{F}_5$)}}
\vspace{-0.1cm}

\finding{4. Third-party apps that do not have the permission to turn
on/off the \nest Cam directly, can do so by modifying the {\em away}
variable ($\mathcal{F}_4$)} The \nest Cam is a home monitoring device, and important for
the users' security. The {\em is\_streaming} variable of the Nest Cam
controls whether the camera is on (\ie streaming) or off, and can only
be written to by an app with the permission {\em Camera r/w}. The \nest
Cam provides a routine as a feature, which allows the camera to be
automatically switched on when the user leaves the home (\ie when the
{\em away} variable of the smart home is set to ``away''), and switched
off when the user returns (\ie when {\em away} is set to ``home'').
Leveraging this routine, third-party apps such as the Belkin Wemo switch
can manipulate the {\em away} field, and indirectly affect the \nest
Cam, without having explicit permission to do so. We tested this ability
with our test app (see Section~\ref{sec:permission-map}) as well, which could indirectly switch the camera {\em on}
and {\em off} at will. This problem has serious consequences; \eg a malicious
test app with the {\em away r/w} permission may set the variable to
``home'' when the user is away to prevent the camera from recording a
burglary. The key problem here is that a {\em low-integrity device/app can
trigger a change in a high-integrity device indirectly}, \ie by modifying a
variable it relies on, which is an instance of the well-known information
flow {\em integrity} problem. Moreover, this is not the only instance of a
high-integrity routine that relies on {\em away}; \eg the Nest x Yale Lock
can lock automatically when the home changes to {\em away} mode~\cite{yalewwn}, 
leakSMART reads the {\em away} state of the home and can notify the user’s
emergency contact when a leak occurs~\cite{leaksmartwwn}. 

\nest has a basic defense to prevent such issues: application design
policies that apply to apps with more than 50 users~\cite{nestreview}.
App developers are required to submit their app for a \review to the
\nest team once the app reaches 50 users, and a violation of the rather
strict and detailed review guidelines can result in the app being
rejected from using the \nest API. One of the review policies (\ie
specifically policy 5.8)  states that \textit{``Products that modify Home/Away
state automatically without user confirmation or direct user action will
be rejected.''~\cite{nestreview}.} \nest users may be vulnerable in spite
of this defense, for two reasons. First, as attacking a smart home is an
attack on a user's personal space, it is feasible to assume that most
attacks that exploit routines will be targeted (\eg to perform
burglaries).  Assuming that the adversary can use social engineering to
get the user to connect a malicious app to their \nest setup, {\em a
targeted attack on a specific user will succeed in spite of the policy},
as the app would be developed solely for the targeted user and hence
will have $<$50 users, and be exempt from the \nest\ \review. Second, it
is unclear how apps are checked against this policy; our next
finding demonstrates a significant omission in \nest's review.

\begin{figure}[t]
    \centering
    \includegraphics[width=2in]{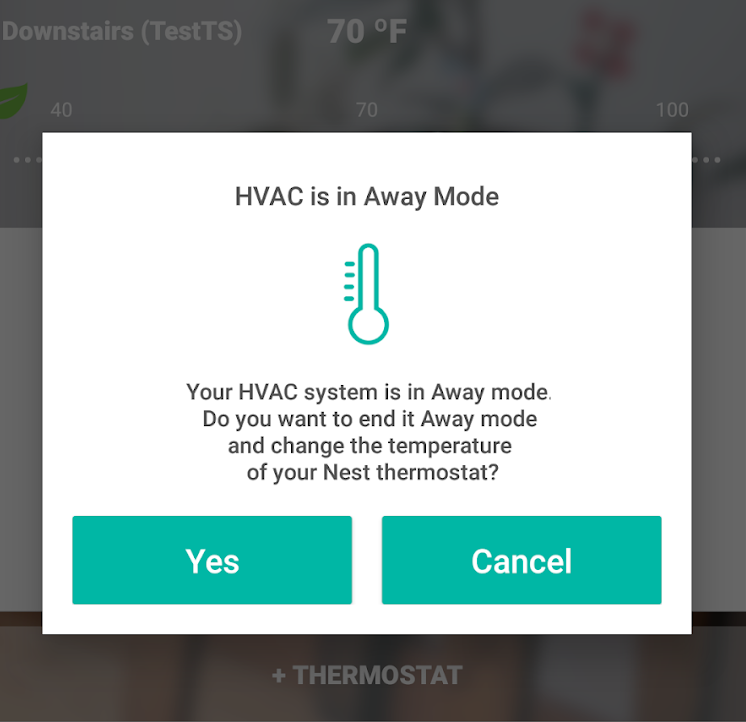} 
    \vspace{-0.7em}
    \caption{{\small The {\em Keen Home} app asks the user to modify
    the thermostat's mode, but in reality, this action leads to the {\em
    entire} smart home being set to ``home'' mode, which affects a
    number of other devices.}}
    \vspace{-1.3em}
\label{fig:keen-home}
\end{figure}

\finding{5. \nest's\ \review policies dictate that the apps must prompt
users before modifying {\em away }, but there is no official constraint on {\em
what} the prompt may display ($\mathcal{F}_5$)} Consider the example in
Figure~\ref{fig:keen-home}, which shows one such prompt by the {\em Keen
Home} app~\cite{keenhomewwn} {\em when the user tries to change the
temperature of the thermostat}. That is, when the user tries to change
the temperature of the thermostat while the {\em away} variable is set
to ``away'', the app requires us to change it to ``home'' before the
thermostat temperature can be changed. This condition is entirely
unnecessary to change the temperature. More importantly, it presents the
prompt to the user in a way that states that the home/away modes are
specific to the HVAC alone.  This is in contrast to the actual
functionality of these modes, in which a change to the {\em away}
variable affects the {\em entire} smart home; \ie we confirmed that the
Nest Cam gets turned off as well once we agree to the prompt. It is
important to note that the {\em Keen Home} app has passed the \nest\
\review, as it has well over 50 users (1K+ downloads on Google
Play~\cite{keenhomeapp}).  Therefore, this case demonstrates that the
\nest\ \review does not consider the contents of the prompt, and a
malicious app may easily misinform the user and make them trigger the
{\em away} variable to the app's advantage. Finally, in
Section~\ref{sec:app-prompts} we demonstrate that this problem of
misinforming the user is not just limited to {\em runtime in-app
prompts} described in this section, but extends
to application-defined {\em install-time permission descriptions}
(\fnumber{6}$\rightarrow$\fnumber{9}).

\section{Security Analysis of \nest Apps}
\label{sec:app}

In this Section, we investigate the privileges of apps developed to be
integrated with \nest.  Unlike prior work~\cite{fjp16}, we not only
report the permissions requested by apps, but also analyze the
information prompts displayed to the user when requesting the
permission.  Additionally, we analyze the rate of SSL misuse by both
general smart home management apps as well as apps integrated with
\nest.  For this section, we do not consider the \hue platform as it has a
limited ecosystem of apps as compared to \nest. We derived two datasets
to perform the analyses that we describe in this section, the \gplayapps
dataset, which contains 650 smart home management apps extracted from
Google Play, and the \nestapps dataset, which includes 39 apps that
integrate into the \nest platform. Our online
appendix~\cite{online_appendix} details our dataset collection
methodologies.

\vspace{-0.9em}
\subsection{Application Permission Descriptions}
\label{sec:app-prompts}
\vspace{-0.1em}

On \nest, developers provide permission descriptions that explain how an
app uses a permission while registering their apps in the \nest
developer console.  These developer-provided descriptions are the {\em
only} direct source of information available to the user to understand
why an app requires a particular permission, \ie\ \nest itself only
provides a short and generic  permission ``title'' phrase that is
displayed to the user along with the developer-defined description (\eg
for {\em Thermostat read}, the \nest phrase is ``See the temperature and
settings on your thermostat(s)'').  Owing to their significant role in
the user's understanding of the permission requirements, we analyze the
{\em correctness} of such developer-defined descriptions relative to the
permissions requested.  

\subsubsection{Analysis Methodology}

As described in Section \ref{sec:characteristics}, upon registering
permissions at the developer console, developers are granted an OAuth
URL that they can direct the user to for obtaining an access token. As a
result, permissions are not encoded in the client mobile app or Web app
(\ie unlike Android), which makes the task of extracting permissions
difficult. However, we observe that the permissions that an app asks for
are {\em always} displayed to the user for approval (\ie when first
connecting an app to their \nest smart home using OAuth). We leverage
this observation to obtain permissions dynamically, \ie by executing
apps to the point of integrating them with our \nest smart home, and
recording the permission prompt displayed for the user's
approval. The procedure is the same for mobile as well as Web apps.  

\subsubsection{\nest App Findings~(\fnumber{6}$\rightarrow$\fnumber{9})}
The two permissions that dominate the permission count are {\em Away
read/write} and {\em Thermostat read/write}, requested by 20
and 24 apps respectively, from the \nestapps dataset.
Our online appendix~\cite{online_appendix} provides the permission count for all
other permissions. Our findings are as follows:

\begin{table*}[t]
\centering
\scriptsize
\vspace{-0.2cm}
\caption{{\small Permission description violations discovered in Works
with \nest apps}}
\vspace{-0.35cm}
\label{tab:descriptions}
\def\arraystretch{1.2}
\begin{tabular}{p{4cm}|p{11.7cm}}
\Xhline{2\arrayrulewidth}
\multicolumn{0}{c|}{\textbf{Application}}   & \multicolumn{1}{c}{\textbf{Incorrect Permission Description}}                                                                                                                                                                                                                                                                                                                                                                                                \\ \Xhline{2\arrayrulewidth}
\multicolumn{2}{c}{\textbf{VC1: Requesting Read/Write instead of Read}} \\ \Xhline{2\arrayrulewidth}

1. Home alerts          & ``{\textbf{thermostat read/write: }}Allows Home alerts to notify you when the Nest temperature exceeds your threshold(s)'' \\ \hline

2. Home alerts          & ``{\textbf{away read/write: }}Allows Home Alerts to notify you when someone is in your home while in away-mode'' \\ \hline

3. MyQ Chamberlain          & ``{\textbf{thermostat read/write: }}Allows Chamberlain to display your Nest Thermostat temperature in the MyQ app'' \\ \hline

4. leakSMART          & ``{\textbf{thermostat read/write: }}Allows leakSMART to show Nest Thermostat room temperature and humidity. New HVAC sensor mode will notify you to shut off your thermostat if a leak is detected in your HVAC system.'' \\ \hline

5. Simplehuman Mirror          & ``{\textbf{Camera+Images read/write: }}Allow your simplehuman sensor mirror pro to capture and recreate the light your Nest Cam sees'' \\ \hline

6. Iris by Lowe's          & ``{\textbf{structure read/write: }}View your Nest Structure names so Iris can help you pair your Nest Structures to the correct Iris Places'' \\ \hline

7. Heatworks model 1          & ``{\textbf{away read/write: }}Allows the Heatworks MODEL 1 to be placed into vacation mode to save on power consumption while you're away'' \\ \hline

8. Feather Controller          & ``{\textbf{Camera+Images read/write: }}Allows Feather to show you your camera and activity images. Additionally, Feather will allow you to request a snapshot.'' \\ \hline

9. Heatworks model 1          & ``{\textbf{thermostat r/w:}} Allows your Heatworks MODEL 1 water heater to go into vacation mode when your home is set to away''                                                                                                                                                                                    \\ \Xhline{2\arrayrulewidth}

\multicolumn{2}{c}{\textbf{VC2: Describing {\em Away} as a property of the thermostat alone, rather than something that affects the entire smart home}} \\ \Xhline{2\arrayrulewidth}

10. Gideon          & ``{\textbf{away read/write: }}Allows Gideon to read and update the Away state of your thermostat'' \\ \hline

11. Muzzley          & ``{\textbf{away read/write: }}Allows Muzzley to read and update the Away state of your thermostat'' \\ \hline                                                                                                                                                                                         

12. Keen home smart vent          & ``{\textbf{away read/write: }}Allows Smart vent to read the state of your Thermostat and change the state from Away to Home''                                                                                                                                                                                                                                                                                                                                                                \\ \Xhline{2\arrayrulewidth}

\multicolumn{2}{c}{\textbf{VC3: Both VC1 and VC2}} \\ \Xhline{2\arrayrulewidth}

13. WeMo          & ``{\textbf{away read/write: }}Allows your WeMo products to turn off when your Nest Thermostat is set to Away and on when set to Home.'' \\ \hline

14. IFTTT thermostat service          & ``{\textbf{thermostat read/write: }}Now you can turn on Nest Thermostat Applets that monitor when you're home, away and when the temperature changes.''
\\ \Xhline{2\arrayrulewidth}

\multicolumn{2}{c}{\textbf{VC4: Descriptions that do not relate to the permission}} \\ \Xhline{2\arrayrulewidth}

15. IFTTT thermostat service          & ``{\textbf{away read/write}}: Now you can set your temperature or turn on the fan with Nest Thermostat Applets on IFTTT''\\ \hline

16. Life360          & ``{\textbf{away read/write:}} We need this permission to automatically turn on/off your nest system''
\\ \Xhline{2\arrayrulewidth}

%
%
%

\end{tabular}
\vspace{-1.5em}
\end{table*}

\finding{6. A significant number of apps provide incorrect permission
descriptions, which may misinform users ($\mathcal{F}_6$)} As shown in
Table~\ref{tab:descriptions}, we found a total of 15 permission
description violations in 13/39 apps from the \nestapps
dataset. We classify these incorrect descriptions into 5 violation
categories (\ie VC1 $\rightarrow$ VC4), based on the specific manner in
which they misinform the user, such as requesting more privileges than 
required for the described need (\eg read/write permissions when only reading is
required), or misrepresenting the effect of the use of the permission
(\eg stating {\em Away} as affecting only the thermostat).  That is,
{\em over 33.33\% of the apps we could integrate have violating
permission descriptions}. 

\begin{figure}[t]
    \centering
    \includegraphics[width=2.5in]{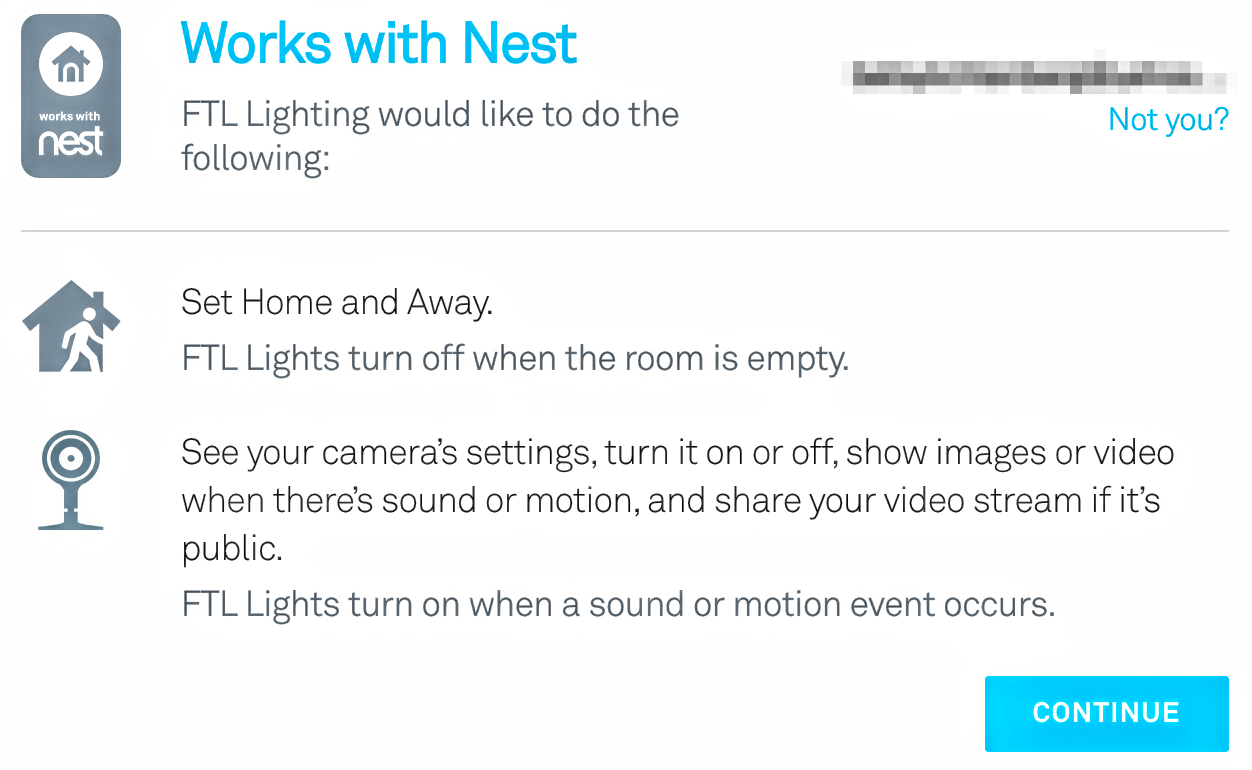}
    \vspace{-1em}
    \caption{{\small An example from the \nest documentation on OAuth
    authorization~\cite{nestoauth} that displays a permission
    description violation (specifically, VC1) for the {\em Away r/w} and
    {\em Camera + images r/w} permissions. The developer's permission
    description indicates that the FTL Lights only need to read data
    store variables, in both cases.
    }} \vspace{-1.5em}
\label{fig:nest_incorrect_desc}
\end{figure} 

\finding{7. In most cases of violations, apps request read/write permissions
instead of read (\fnumber{7})} In 9 cases, apps request the more privileged {\em read/write}
version of the permission, when they should have clearly requested the
{\em read} version, as per their permission description (\ie VC1 in
Table~\ref{tab:descriptions}). For example, consider the ``MyQ
Chamberlain'' app (Table~\ref{tab:descriptions}, entry 3), which asks
for the {\em thermostat read/write} permission, but whose description only
suggests the need for the {\em thermostat read} permission, \ie ``Allows
Chamberlain to display your Nest Thermostat temperature in the MyQ
app''. More importantly, a majority of the violations of this kind occur
for the {\em Away read/write} and {\em Camera+Images read/write} permissions, which
may have serious consequences if these overprivileged apps are
compromised, \ie as {\em Away read/write} regulates control over indicating whether a user is at home or out of the house, and {\em Camera+Images read/write}  may allow apps to turn off
the \nest cam via the {\em is\_streaming} variable. 
These violations exist in spite of \nest guidelines
that mention the following as a {\em Key Point}: \textit{``Choose `read'
permissions when your product needs to check status. Choose
`read/write' permissions to get status checks and to write data
values.''}~\cite{nestperms}. Finally, we found that the {\em \nest
documentation may itself have incorrect instructions}, 
\eg the \nest's documentation on OAuth 2.0
authentication~\cite{nestoauth} shows an example permission prompt 
that incorrectly requests the {\em
Away read/write} permission while only needing read access, \ie with the
description ``FTL Lights turn off when the room is empty'', as shown in
the Figure~\ref{fig:nest_incorrect_desc}.

\finding{8. The \nest\ \review is insufficient when it comes to
reviewing the correctness of permission descriptions and requests by
apps (\fnumber{8})} The \nest\ \review suggests the following two
rules, violating which may cause apps to be rejected: (1) \textit{``3.3.
Products with names, descriptions, or permissions not relevant to the
functionality of the product''}, and (2) \textit{``3.5. Products that have
permissions that don't match the functionality offered by the
products''}~\cite{nestreview}. Our findings demonstrate that the 16
violations discovered violate either one or both of these rules (\eg by
requesting read/write permissions, when the app only requires read).
The fact that the apps are still available suggests that the \nest~
\review may not be rigorously enforced, and as a result, may be
insufficient in protecting the attacks discovered in
Section~\ref{sec:platform-routines}.

\finding{9. Apps often incorrectly describe the {\em Away} field as a
local field of the \nest thermostat, which is misleading
(\fnumber{9})} One example of this kind (VC2 in
Table~\ref{tab:descriptions}) is the {\em Keen Home} app described in
Section~\ref{sec:platform-routines} (Table~\ref{tab:descriptions}.
entry 12), which states that it needs {\em Away read/write} in order to \textit{``Allow Smart
vent to read the state of your Thermostat and change the state from Away
to Home''}. As a result, {\em Keen Home} misrepresents the effect and
significance of writing to the {\em Away} field, by making it seem like
{\em Away} is a variable of the thermostat, instead of a field that
affects numerous devices in the entire smart home.  Gideon and Muzzley
(entries 10 and 11 in Table~\ref{tab:descriptions}) present a similar
anomaly.
Our hypothesis is that such violations occur
because \nest originally started as a smart thermostat that gradually
evolved into a smart home platform. 
Finally, in addition to misleading descriptions classified as VC1 and VC2, we
discovered apps whose permission descriptions did not relate to the
permissions requested at all (VC4), and apps whose descriptions
satisfied both VC1 and VC2 (\ie VC3 in Table~\ref{tab:descriptions}).

The accuracy of permission descriptions is important, as the user has no
other source of information upon which to base their decision to trust
an app.  \nest recognizes this, and hence, makes permissions and
descriptions a part of its \review. The discovery of inaccurate
descriptions not only demonstrates that apps may be overprivileged, but
also that \nest's design review process is incomplete, as it puts all
its importance on getting the user's consent via permission prompts (\eg
in Findings 5$\rightarrow$9), but not on what information is actually
shown.

\vspace{-0.22cm}
\subsection{Application SSL Use}
\label{sec:ssl}
\vspace{-0.08cm}

The previous section demonstrated that smart home apps may be
overprivileged in spite of a dedicated \review. An adversary may be able
to compromise the smart home by exploiting vulnerabilities in such
overprivileged apps. As a result, we decided to empirically derive 
an estimate of how vulnerable smart home apps are, in terms of
their use of SSL APIs, which form an important portion of the apps'
attack surface.

We used two datasets for this experiment, \ie the \gplayapps dataset
consisting of 650 generic smart home (Android) apps crawled from Google
Play, and an extended version of the \nestapps dataset, \ie the
\nestappsext dataset, which consists of 111 Android apps built for Works
with \nest devices (\ie including the ones for which we do not possess
devices).  We analyzed each app from both the datasets using
MalloDroid~\cite{fhm+12}, to discover common SSL flaws.

\finding{10. A significant percentage of general smart home management apps, as well as apps that connect to \nest have serious SSL
vulnerabilities ($\mathcal{F}_{10}$)} 20.61\% (\ie 134/650) of the smart home apps from the
\gplayapps dataset, and 19.82\% (\ie 22/111) apps from the \nestappsext
dataset, have at least one SSL violation as flagged by MalloDroid.
Specifically, in the \nestappsext dataset, the most common cause of an
SSL vulnerability is a broken {\em TrustManager} that accepts {\em all
certificates} (\ie 20 violations), followed by a broken {\em
HostNameVerifier} that does not verify the hostname of a valid
certificate (\ie 11 violations). What is particularly worrisome is that
apps such as {\em MyQ Chamberlain} and {\em Wemo} have multiple SSL
vulnerabilities as well as the {\em Away read/write} permission, which makes
their compromise especially dangerous. Prior work has demonstrated that
such vulnerabilities can be dynamically exploited (\eg via a
Man-in-the-Middle proxy)~\cite{fhm+12,rsb+15}, and we use similar
approaches to demonstrate an end-to-end attack on the \nest security
camera, using one of the SSL vulnerabilities discovered from this
analysis.

\section{Lateral Privilege Escalation} \label{sec:attack}

While our findings from the previous sections are individually
significant, we demonstrate that they can be combined to form an instance of a lateral
privilege escalation attack~\cite{lateral-esc}, in the context of smart
homes. That is, we demonstrate how {\em an adversary can compromise one
product (device/app) integrated into a smart home, and escalate
privileges to perform protected operations on another product,
leveraging routines configured via the centralized data store.} 

This attack is interesting in the context of smart homes, because of two
core assumptions that it relies on (1) low-integrity (or non-security)
smart home products may be easier to directly compromise than
high-integrity devices such as the \nest Cam (\ie none of
the SSL vulnerabilities in \fnumber{10} were in security-sensitive
apps), and (2) while low-integrity devices may not be able to directly
modify the state of high-integrity devices (\fnumber{1}), they may be
able to indirectly do so via {\em automated routines} triggered by
global smart home variables (\fnumber{4}). (3) Moreover, since the
low-integrity device is not being intentionally malicious, but is
compromised, the \review process would not be useful, even if it was
effective (which it is not, as demonstrated by
\fnumber{5}$\rightarrow$\fnumber{9}). This last point distinguishes a
lateral privilege escalation from actions of malicious apps that trigger
routines (\eg the ``fake alarm attack'' discussed in prior
work~\cite{fjp16}). These conditions make lateral privilege escalation
particularly interesting in the context of smart home
platforms, and especially, DSB platforms such as SmartThings and \nest.

\myparagraph{Attack Scenario and Threat Model} We consider
a common man-in-the-middle (MiTM) scenario, similar to the
SSL-exploitation scenarios that motivate prior
work~\cite{fhm+12,rsb+15}. Consider Alice, a smart home user who has
configured a security camera to record when she is away (\ie using the
{\em away} variable in the centralized data store). Bob is an
acquaintance (\eg a disgruntled employee or an ex-boyfriend) whose
motive is to steal a valuable from Alice's house without being recorded
by the camera. We assume that Bob also knows that Alice uses a smart
switch in her home, and controls it via its app, which is integrated
with Alice's smart home.  Bob follows Alice, and connects to the same
public network as her (\eg a coffee shop, common workplace), sniffs the
access token sent by the switch's app to its server using a known SSL
vulnerability in the app, and then uses the token to directly control
the {\em away} variable. Setting the {\em away} to ``home'' confuses the
security camera into thinking that Alice is at home, and it stops
recording. Bob can now burglarize the house without being
recorded.  

\myparagraph{The Attack} The example scenario described previously can
be executed on a \nest smart home, using the \nest Cam and the TP Link
Kasa switch (and the accompanying Kasa app).  We compromise the SSL
connection of Kasa app, which was found to contain a broken SSL
TrustManager in our analysis described in Section~\ref{sec:app}. We
choose Kasa app as it requests the sensitive {\em Away read/write}
permission, and has a sizable user base (1M+ downloads on Google
Play~\cite{kasastats}). It is interesting to note that the Kasa app has
also passed the \nest product review process and is advertised on the
Works with \nest website~\cite{kasaworkswithnest}, but can still be
leveraged to perform an attack. We use {\em bettercap}~\cite{bettercap}
as a MiTM proxy to intercept and modify unencrypted data. Additionally,
as described in the attack scenario, we assume that {\sf (1)} the victim's
\nest smart home has the \nest Cam and the Kasa switch installed, {\sf
(2)}
the popular routine which triggers the \nest Cam to stop recording when
the user is home is enabled, and {\sf (3)} the user connects her
smartphone to a network to which the attacker has access (e.g., coffee
shop, office), which is a common assumption when exploiting
SSL-misuse~\cite{fhm+12,rsb+15}.  

\begin{lstlisting}[basicstyle=\ttfamily\scriptsize,float, caption={The Kasa app's unencrypted GET request.},label=lst:get-request,emph={accountToken,email},emphstyle=\bfseries]
{"data":{"uri":"com.tplinkra.iot.authentication.impl.RetrieveAccountSettingRequest"},
	"iotContext":
		{"userContext":{"accountToken":"<anonymized alphanumeric token>",
		"app":{"appType":"Kasa_Android"},
		"email":"<anonymized>",
		"terminalId":"<anonymized>"}}, ...
\end{lstlisting}

The attack proceeds as follows:
{\sf (1)} The user utilizes the Kasa app to control the switch,
while the user's mobile device is connected to public network. 
{\sf (2)} The attacker uses a MiTM proxy to intercept Kasa app's
attempt to contact its own server, and supplies the attacker's
certificate to the app during the SSL handshake, which is accepted by
the Kasa app due to the faulty TrustManager. 
{\sf (3)} The Kasa app then sends an authorization token (see
Listing~\ref{lst:get-request}) to the MiTM proxy
(\ie assuming it is the authenticated server), which is stolen
by the attacker.  This token authorizes a particular client app to send
commands to the TP Link server.
{\sf (4)} Using the stolen token, the attacker instructs the TP Link server to
set the smart home's {\em away} variable to the value ``home'', while the user
is actually ``away''. This action is possible as the TP Link server (\ie Web
app) has the {\em -Away read/write} permission for the user's \nest smart home.
{\sf (5)} This triggers the routine in the \nest Cam, which stops
recording.

In sum, the attacker compromises a security-insensitive (i.e.,
low-integrity) product in the system, and uses it along with a routine
to escalate privileges, \ie to modify the state of a security-sensitive
(i.e., high-integrity) product. It should be noted that while this is
one verified instance of a lateral privilege escalation attack on DSB
smart home platforms, given the broad attack surface indicated by our
findings, it is likely that similar undiscovered
attacks exist.

\section{Lessons}
\label{sec:lessons}

Our findings 
($\mathcal{F}_1$)$\rightarrow$($\mathcal{F}_{10}$) demonstrate numerous
gaps in the security of smart home platforms that implement routines
using centralized data stores. Moreover, while many of the findings may
apply to platforms such as SmartThings as well, their implications are
more serious on \nest, as the user does not have a centralized
perspective of the routines programmed into the smart home. We now
distill the core lessons from our findings, which motivate significant
changes in modern platforms such as \nest.

\myparagraph{Lesson 1 }{ \em Seamless automation must be accompanied by
strong integrity guarantees.}
It is important to note that the attack described in
Section~\ref{sec:attack} may not be addressed by fixing the problem of
overprivilege or via product reviews, since none of the components of
the attack are overprivileged (\ie including TP Link Kasa), and our
findings demonstrate that the \nest\ \review is insufficient
(\fnumber{5}$\rightarrow$\fnumber{9}). The attack was enabled due to the
integrity-agnostic execution of routines in \nest (\fnumber{4}). To
mitigate such attacks, platforms such as \nest need information flow
control (IFC) enforcement that ensures strong integrity
guarantees~\cite{bib77}, and future work may explore the complex
challenges of (1) specifying integrity labels for a diverse set of user
devices and (2) enforcing integrity constraints without sacrificing
automation. Moreover, as third-party devices are integrated into the
data store, future work may also explore the use of {\em decentralized}
information flow control (DIFC) to allow devices to manage the integrity
of their own objects~\cite{ml97,kyb+07,zbkm06}. The introduction
of {\em tiered-trust domains} in \nest (\ie via Weave) offers an
encouraging start to the incorporation of integrity guarantees into
smart home platforms~\cite{nestweave}.

\myparagraph{Lesson 2} {\em \nest Product Reviews would benefit from at
least light-weight static analysis.}
Our findings demonstrate numerous violations of the \nest design
policies that should have been discovered during the \review.  Moreover,
the review guidelines also state that products that do not securely
transmit tokens will be rejected~\cite{nestreview}, but our simple
static analysis using MalloDroid discovered numerous SSL vulnerabilities
in \nest apps (\fnumber{10}), of which one can be exploited
(Section~\ref{sec:attack}). We recommend the integration of light-weight
tools such as MalloDroid in the review process.

\myparagraph{Lesson 3} {\em The security of the smart home 
indirectly depends on the smart phone (apps).}  Smartphone apps have
been known to be susceptible to SSL misuse~\cite{fhm+12}, among
other security issues (\eg unprotected interfaces~\cite{cfgw11}).
Thus, unprotected smartphone clients for smart home devices may enable
the attacker to gain access to the smart home, and launch further
attacks, as demonstrated in Section~\ref{sec:attack}. Ensuring the
security of smart phone apps is a hard problem, but future work may
triage smartphone apps for security analyses based on the volume of
smart home devices/platforms they integrate with, thereby, improving the
apps that offer the widest possible attack surface to the adversary.

\myparagraph{Lesson 4} {\em Popular but simpler
platforms need urgent attention.} The startling gaps in the access
control of \hue demonstrate that the access control of other simple
(\ie homogeneous) platforms may benefit from a similar holistic security
analysis (\fnumber{2}, \fnumber{3}).

\section{Vulnerability Reporting}
\label{sec:reporting}
We have reported the discovered vulnerabilities to Philips
(\fnumber{2},~\fnumber{3}), Google
(\fnumber{1},~\fnumber{4}$\rightarrow$\fnumber{9}), and TP Link
(\fnumber{10}), and have received confirmations from all the vendors. TP
Link has since fixed the SSL flaw in the latest version of the app.
Philips \hue is currently  analyzing third party apps for the specific
behavior discussed in this paper, and will eventually roll out a fix to
their access control policy. We have also provided recommendations to
Google on improving the safety of routines, which is a design challenge
that may be hard to immediately address.

\section{Threats to Validity}
\label{sec:threats}

\myparagraph{1. SSL MiTM for different Android versions} Our attack
described in Section~\ref{sec:attack} has been tested and is fully
functional on a Nexus 7 device running Android version 4.4.2. However,
we have recently observed that the MiTM proxy is blocked when
intercepting connections from a Pixel 2 device running the latest
version of Android (\ie 8.1.0). Our hypothesis is that the TP Link Kasa
app changes its SSL API use based on the Android API version, and we are
currently working on locating at what Android version (\ie between 4.4.2 and 8.1.0) the SSL component of our described attack no longer functions. However, this caveat does not change the fact that our attack is feasible under certain settings, or that third-party Android apps may often have exploitable SSL verification
vulnerabilities~\cite{fhm+12,od15,rsb+15}. It is important
to note that the SSL compromise is a well-studied engineering challenge,
and is not the focal point of the lateral privilege escalation exploit
we describe, which occurs primarily because of routines implemented using
shared global variables in \nest (\fnumber{4}$\rightarrow$\fnumber{6}).

\myparagraph{2. Number of devices and apps} For the analysis in
Section~\ref{sec:app}, our set of 9 devices (\ie 7 real and 2 virtual)
allowed us to integrate a set of 39 apps into our \nest platform (\ie
the \nestapps dataset), out of the around 130 ``Works with'' \nest apps
we found. Therefore, while we cannot say that our findings
(\fnumber{6}$\rightarrow$\fnumber{9}) generalize to all the apps compatible
with Nest, we can certainly say that they are valid for a significant
minority (\ie over 27\%). 

\myparagraph{3. Local and Remote exploits of \hue} Our exploits for the Philips Hue platform
demonstrated in Section~\ref{sec:permission-map} (\fnumber{2} and
\fnumber{3}) can be executed from an app operating on the same local
network as the \hue bridge. This is feasible, as the
attacker-controlled app simply needs to be on the same network (\ie not
even on the victim's device). The vulnerabilities we describe may also
be remotely exploited, as access control enforcement remains the
same for remote access.  

\section{Related Work}
\label{sec:relwork}

Smart home platforms are an extension of the new modern OS paradigm,
the security problems in smart home platforms are similar to
prior modern OSes (\eg application over-privilege, incorrect platform
enforcement).  As a result, some of the same techniques may be applied
in detecting such problems. For instance, in a manner similar to Felt et
al.'s seminal evaluation of Android permission
enforcement~\cite{fch+11}, our work uses automated testing to derive
permission maps and compares the maps to the platform documentation. We
also leverage lessons from prior work on SSL
misuse~\cite{fhm+12,od15,rsb+15,ssl+14} to perform the SSL
Analysis (Section~\ref{sec:ssl}) and the MiTM exploit
(Section~\ref{sec:attack}). The lack of transitivity in access control
that we observe is similar to prior observations on
Android~\cite{fwm+11,cfgw11,ne13,naej16,dsp+11}. However, the
implications of intransitive enforcement are different in the smart home
space, and, to our knowledge, some of the key analyses performed in this
paper is novel across modern OS research (\eg exploitation of home
automation routines and the ineffectiveness of \nest's product review).
The novelty of this paper is rooted in using lessons learned from prior
research in modern OS and application security to identify problems in
popular but under-evaluated platforms such as \nest and \hue, and
moreover, in demonstrating the potential misuse of home automation {\em
routines} for performing lateral privilege escalation. 

In the area of smart home security, the investigation by Fernandez et
al.~\cite{fjp16} into the SmartThings platforms and its apps is
highly related to the study presented in this paper.  
However, our work exhibits key differences. For instance, the platforms explored in this paper
(\ie\ \nest and \hue) are popular, and have key differences relative to
SmartThings (Section~\ref{sec:characteristics}). Moreover, while
Fernandez et al.  focus on application overprivilege, this work studies
the utility and security of routines, and leverages routines to
demonstrate the first instance of lateral privilege escalation on smart
home platforms.  Our analysis of permission text artifacts,
\review-based defense in \nest, and SSL-misuse in apps leads to
novel findings that facilitate the end-to-end attack. Finally, we
demonstrate that simpler platforms (\ie~ \hue) fail to provide
bare-minimum protections. 

Aside from this closely-related work, prior work has demonstrated direct
attacks on smart home platforms and applications. For instance, Sukhvir
et al. attack the communication and authentication protocols in \hue and
Wemo~\cite{nsg+14}, Sivaraman et al. attack the home's firewall using a
malicious device on the network~\cite{sceb16}, and a Veracode study
demonstrated  issues in a range of products such as the MyQ Garage
System and Wink Relay~\cite{veracode}. Our work performs a holistic
security evaluation of the access control enforcement in DSB platforms
(\ie\ \nest and \hue) and their applications, and is complementary to
such per-device security analysis.  

Prior work has also analyzed the security of trigger-action programs.
Surbatovich et al.~\cite{sab+17} analyzed the security and privacy risks
associated with IFTTT recipes, which are trigger-action programs similar
to routines. The key difference is that Surbatovich et al. examines the
safety of individual recipes, while our work explores routines that may
be safe on their own (\eg when {\em home}, turn off the \nest Cam), but
which may be used as gadgets by attackers to attack a high-integrity
device from a low-integrity device.

In a similar vein, Celik et al.~\cite{cmt18} presented Soteria, a static
analysis system that detects side-effects of concurrent execution of
Samsung's smart apps. The problem explored in our paper is broadly
similar to Celik et al.'s work, \ie both papers explore problems that
arise due to the lack of transitive access control in smart homes.
While the techniques that underlie Soteria have advanced the state of
the art for analyzing smart home products, our paper exhibits two key
differences that demonstrate the novelty of our analysis. First, Soteria
does not aim to address the adversarial use of routines as mechanisms to
perform a lateral privilege escalation. As a result, it would not detect
the attack discussed in Section~\ref{sec:attack}, since the precondition
for the attack is not a routine (\ie it is the exploitation of SSL
vulnerability in the Kasa app, which allows us to steal the
authorization token and  misuse the {\em away} permission allocated to
Kasa).  Second, this paper is novel in its analysis of runtime prompts and
permission descriptions on home automation platforms, and uncovers
problems in how users are informed of specific sensitive automation
actions (\fnumber{8}$\rightarrow$\fnumber{9}), and how the permissions
that enable such actions (\fnumber{5}) are described. 

Finally, prior work has proposed novel access control enhancements,
which may alleviate some of the concerns raised in this paper.
ProvThings~\cite{whbg18} provides provenance information that may allow
the user to piece together evidence of some of the attacks
described in this paper, but does not prevent the attacks themselves. On
the contrary, ContextIoT~\cite{jcw+17} provides users with runtime
prompts describing the context of sensitive data accesses, which may
alert users to unintended execution of routines (\fnumber{4}), {\em at
the cost of reducing automation}. 
Further, SmartAuth~\cite{tzl+17} analyzes the consistency of application
descriptions with code, and may benefit the \nest\ \review in
determining the correctness of permission descriptions.

\vspace{-0.2cm}
\section{Conclusion}
\label{sec:conc}

Smart home platforms and devices operate in the
users' physical space, hence, evaluating their security is critical. 
This paper evaluates the security of two such platforms, \nest and \hue, that
implement home automation {\em routines} via centralized data stores.
We systematically analyze the limitations of the access control enforced by
\nest and \hue, the exploitability of routines in \nest, the robustness
of \nest's \review, and the security of third-party apps that integrate
with \nest.
Our analysis demonstrates {\sf ten} impactful findings, which
we leverage to perform an end-to-end 
lateral privilege escalation attack in the context of the smart home. Our
findings motivate more systematic and design-level
defenses against attacks on the integrity of the users' smart home.

\bibliographystyle{ACM-Reference-Format}
\bibliography{ms}

\end{document}